\documentclass{article}
\usepackage[utf8]{inputenc}
\usepackage[T1]{fontenc}
\usepackage[english]{babel}
\usepackage{graphicx}
\usepackage{floatflt}
\usepackage{amsthm}
\usepackage{amsmath}
\usepackage{amssymb} 
\usepackage[usenames, dvipsnames]{color}
\usepackage{mathrsfs}
\usepackage{geometry}  
\usepackage{makeidx}
\usepackage{physics}
\usepackage{mathtools}
\usepackage{authblk}
\usepackage{lineno}
\usepackage[numbers,sort&compress]{natbib}
\usepackage{csquotes}
\usepackage{chngcntr}
\usepackage{setspace}
\usepackage[labelfont=bf]{caption}
\usepackage{comment}

\DeclareUnicodeCharacter{03BC}{$\mu$}
\DeclareUnicodeCharacter{2009}{ }
\DeclareUnicodeCharacter{03BB}{\ensuremath{\lambda}}
\DeclareUnicodeCharacter{223C}{\ensuremath{\sim}}

\title{Quantum Cascade Lasers as Broadband Sources via Strong RF Modulation}

\author[1,*]{A. Cargioli}
\author[1]{D. Piciocchi} 
\author[1]{M. Bertrand}
\author[2]{R. Maulini}
\author[2]{S. Blaser}
\author[2]{T. Gresch}
\author[2]{A. Muller}
\author[1]{G. Scalari}
\author[1]{J. Faist}

\affil[1]{Institute for Quantum Electronics, ETH Zurich, CH-8093 Zurich, Switzerland}
\affil[2]{Alpes Lasers, Avenue des Pâquiers 1, 2072 St Blaise, Switzerland}

\affil[*]{Contact Email: acargioli@phys.ethz.ch}

\date{}
\begin{document}

\modulolinenumbers[1]

\renewcommand{\figurename}{Figure}
\def\equationautorefname#1#2\null{Eq.#1(#2\null)}
\maketitle
\numberwithin{equation}{section}

\begin{abstract}
In this work, we demonstrate that in a regime of strong modulation, by generating pulses of the length of the order of a few cavity lifetimes (hundreds of ps), a broadband quantum cascade laser can be driven to lase on a bandwidth (250cm$^{-1}$) limited by the gain. In addition, the amplitude noise of the radiation was shown to be limited by the detector. A laser linewidth study has been performed under different operating conditions finding values spanning from 20MHz to 800MHz, indicating a trade-off between emission bandwidth, amplitude stability and coherence.

\end{abstract}

\setcounter{section}{1}
\counterwithout{equation}{section}

\newpage

\section*{Introduction}

The Mid-Infrared (Mid-IR) spectral region is of particular interest in many different fields, spanning from environmental monitoring \cite{willer_near-_2006} to biomedical diagnostics \cite{wang_application_2008}. The possibility of performing non-invasive, non-destructive and fast spectroscopy is extremely appealing and goes hand in hand with the necessity of having bright, stable and broadband sources. Among the many possible Mid-IR sources, thermal emitters provide complete coverage of the Mid-IR spectral range, albeit at the cost of low emission power spectral densities\cite{jung_next-generation_2017} which is not well suited for microscopy and high-resolution spectroscopy. Recent developments on superlattice light emitting diodes \cite{ricker_broadband_2017} attempted to solve the power emission problems, showing an effective thermal emission between 3 and 5 $\mu$m with radiances of over 1 W/cm$^2$. However, due to their need to operate at cryogenic temperatures, LEDs remain  not particularly suited for Mid-IR applications due to the intrinsically lower spontaneous emission rate. Another approach to generate broadband emitters extending in the Mid-IR is exploiting non-linearity in various dielectric media to achieve supercontinuum generation \cite{yu_experimental_2016, zorin_advances_2022} which allows a broadband spectral coverage but with the need of high power pulsed lasers, therefore adding complexity to the measurement system. Furthermore, the amplitude noise stability of these systems is difficult to control \cite{corwin_fundamental_2003, klimczak_direct_2016} which strongly impacts the possible spectroscopy applications. Another well established state of the art platform is represented by Quantum Cascade Lasers (QCLs)\cite{faist_quantum_1994} and Interband Cascade Lasers (ICLs) \cite{lin_type-ii_nodate}. They are electrically driven, they have an extremely versatile emission wavelength tunability through bandgap engineering with the ability of creating a system with atom-like energy transitions, and it has been demonstrated that they can produce optical frequency combs (FC) \cite{hugi_mid-infrared_2012, faist_quantum_2016, sterczewski_interband_2021}. The possibility of producing on chip, electrically pumped self-starting frequency combs opened the way to dual comb spectroscopy using both QCLs \cite{villares_dual-comb_2014, gianella_high-resolution_2020} and ICLs \cite{sterczewski_mid-infrared_2020}, which makes extremely fast measurements in a very compact setup possible. To this day, the main limitation of these platforms is given by the emission bandwidth of the devices, that can reach a maximum of about 90 cm$^{-1}$ for ICLs \cite{feng_passively_2020} and 100 cm$^{-1}$ for QCLs \cite{singleton_evidence_2018, heckelmann_quantum_2023}, limiting broadband spectroscopy applications. Concerning QCLs, broad gain active regions based on heterogeneous integration have been demonstrated \cite{gmachl_ultra-broadband_2002, bandyopadhyay_high_2014}, but they are typically used for tunable external cavity QCLs (EC-QCL) \cite{hugi_external_2010}. The interplay of dispersion and Kerr nonlinearity, which are extremely difficult to engineer on broadband active regions, have a profound impact on the stability of the system \cite{villares_quantum_2015,opacak_theory_2019,beiser_engineering_2021}, in particular the emission modes can interact through the medium without ever finding a stable state, thus creating a very high amplitude and frequency noise. For this reason these kind of structures have never proven to be able to support stable frequency comb generation, preventing, on one hand, to be used in Dual Comb Spectroscopy for lack of coherence, and more in general for spectroscopy application for their intrinsically high amplitude noise. Because these ultra broadand active regions do not naturally support stable broadband comb operation, we use them in a way that prevents the Four Wave Mixing (FWM) to act with a prominent effect, which typically happens on timescales of hundreds of cavity roundtrips ($\sim 100$ns) \cite{singleton_combs_2021}. To this end, by using a low frequency, high power radio frequency (RF) modulation scheme, we turn on and off the laser as soon as the power reaches its peak, i.e. on a timescale of a few photon cavity lifetimes (hundreds of ps). Therefore, during the operation time where the device has enough gain to lase, there is enough time for the self-gain saturation of each mode to be relevant, allowing the multimode operation which can cover the full gain bandwidth, but at the same time the cross gain, which is typically smaller \cite{khurgin_coherent_2014}, does not play a significant role. Therefore the non-linear dynamic, which would act on the phase-locking of the modes, is too slow to be able to create instabilities thus improving the mode amplitude stability in a device that would naturally operate in an extremely noisy frequency-comb like emission regime. Note that this is fundamentally different from the strong modulation of the laser at the cavity round trip frequency as addressed in previous works \cite{hillbrand_coherent_2019, schneider_controlling_2021,st-jean_injection_2014} which enables the stabilization and broadening of a pre-existing self-starting frequency comb.

\section*{Results and Discussion}

\begin{figure}[htb!]
  \centering
  \includegraphics[width=\textwidth]{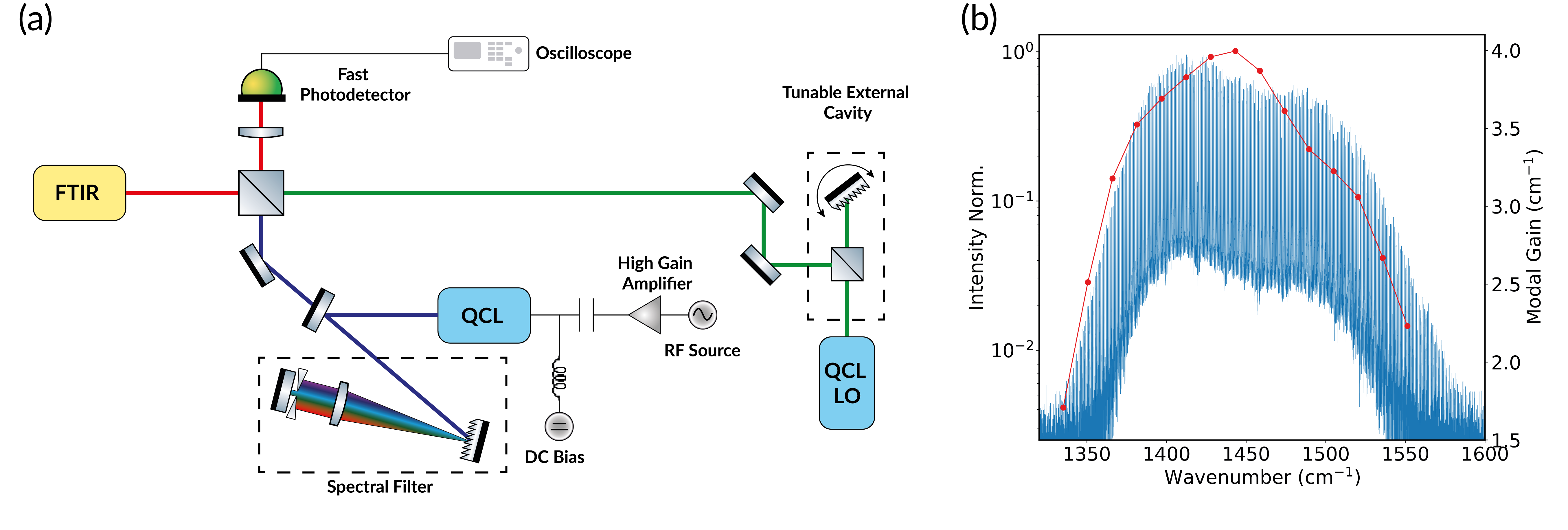}
  \caption{\textbf{a)} Schematic of the experimental setup. \textbf{b)} Spectrum of the emitted light under strong modulation ($f_M = 800$ MHz, $P_M = 29$ dBm) with a current bias of 350 mA (blue curve), and measured sub-threshold modal gain (red curve).}
  \label{fig:setup}
\end{figure}

The device under study has a multistack active region structure \cite{hugi_external_2009,bidaux_measurements_2015} processed through a buried heterostructure technique \cite{beck_continuous_2002} emitting around 7 $\mu$m ($\sim 1430$cm$^{-1}$). To operate the device, we use a bias tee to superimpose the RF modulation to the DC current. The RF signal is obtained by amplifying a low frequency signal (between 0.5 and 1 GHz) through a high gain amplifier (49 dBm saturation power). In order to measure the spectral properties (emission spectrum and coherence), the emitted light is sent through a spectral filter operating in reflection,  where the different frequency components dispersed by a diffraction grating are selected using a slit. By changing the slit aperture is possible to select the spectrum from its full bandwidth down to only few modes. After the filter, the light can either be sent to an FTIR or mixed with a tunable EC-QCL to perform heterodyne detection as shown in the schematic of Fig.\ref{fig:setup}(a). The typical emission spectrum under strong modulation presents a regular and smooth envelope, covering a broad spectral region up to 250 cm$^{-1}$. An example is reported in Fig.\ref{fig:setup}(b) for a modulation frequency $f_M = 800$ MHz and a power $P_M = 29$ dBm, where the spectrum is compared with the modal gain profile, showing  that the device can be operated close to the bandwidth of its gain. The latter has been measured by the Hakki-Paoli technique using an FTIR measurement \cite{hofstetter_measurement_1999}. 

\begin{figure}[htb!]
  \centering
  \includegraphics[width=.9\textwidth]{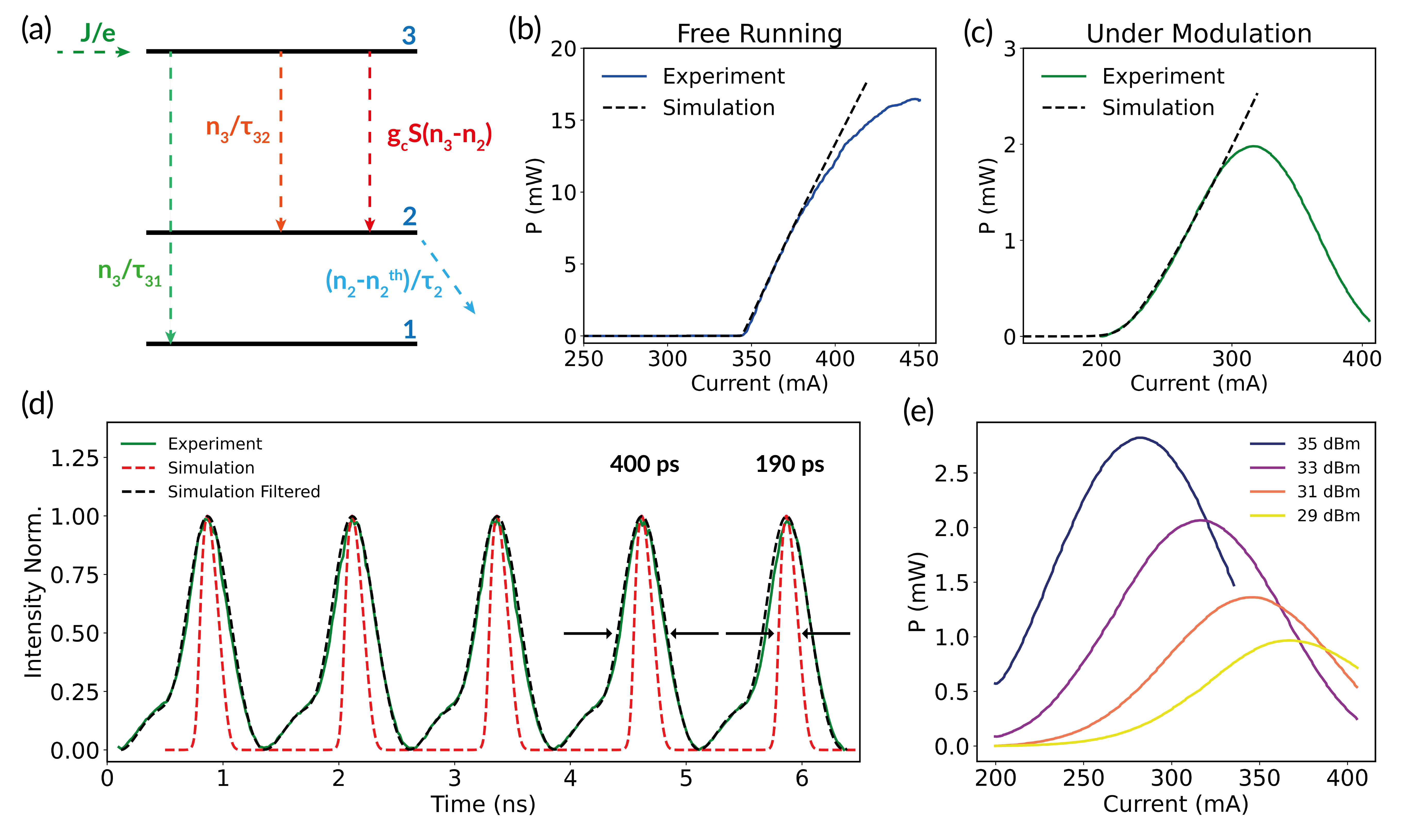}
  \caption{\textbf{(a)} Schematic representation of the 3-level system used in the rate equation model to predict the light-current curves and the time dynamics of the QCL. \textbf{(b)} Light-Current curve in the free running case. Both the measurement (solid line) and the simulated (dashed line, $\sigma = 2.5$) curves are reported. \textbf{(c)} Light-Current curve in the strong modulation case ($P_M = 33$ dBm and $f_M = 800$ MHz). Both the measurement (solid line) and the simulated (dashed line, $J_{M}^{0}  = 1.07$ kA/cm$^2$, $\sigma = 9.3$) curves are reported. \textbf{(d)} Measured (solid green line), simulated (dashed red line) and simulated with the detector filtering (dashed red line) time resolved pulses of the device under strong modulation ($P_M = 33$ dBm, $f_M = 800$ MHz, current bias of 267 mA). The pulse-width measured from the detector trace is of about 400 ps, while the estimated real pulse width is of about 190ps. \textbf{(e)} Light-Current curves under strong modulation at $f_M = 800$ MHz for different modulation powers.  }
  \label{fig:LI-pulses}
\end{figure}

By driving the system with such a strong modulation, since the gain is never clamped, we drastically change its behavior with respect to the free-running condition. A a 3-level rate equation model (Fig.\ref{fig:LI-pulses}(a)) \cite{faist_quantum_2013} has been implemented to describe the laser emission under strong modulation (see Supp. Mat. for further details). This model remains very simplified, as it considers only one single mode and does take neither gain saturation nor thermal effects into account. However, it captures reasonably well the laser dynamics if we limit ourselves to the first part of the light-current characteristic. The laser parameters (see Tab.\ref{tab:rate} in Supp. Mat.) are first estimated fitting the LIV curve measured in CW operation, reported in Fig.\ref{fig:LI-pulses}(b). Since the model does not take into account any thermal behavior, we expect to overestimate the slope efficiency. Therefore, to fit the curves, we introduce a power scaling parameter $\sigma$ which empirically takes into account the heating dependence (see S\ref{SI:scaling} in Supp. Mat.). The simulated curve follows well the trend of the measured one up to 380 mA, where the gain saturation starts playing an important role. In the case of the strongly modulated system, a current modulation term $J_M(t) = J_{M}^{0} \sin (\omega t)$ is added to the constant current term $J_0$ and introduced in the rate equations. $J_{M}^{0} $ is the current density modulation amplitude and $\omega$ is the modulation frequency . The comparison between a strongly modulated light-current curve for a modulation frequency of $f_M = 800$ MHz and power $P_M = 33$ dBm and the relative simulated curve is reported in In Fig.\ref{fig:LI-pulses}(c). From the RF power, it is possible to estimate a current amplitude sweep of about 255 mA. The two curves match up to a DC current bias of 290 mA where, as in the previous case, gain saturation start playing an important role as the peak current reaches now 540mA. These parameters are also used to extract the laser dynamics at a DC current bias of 267 mA, which is directly compared with a measured trace in Fig.\ref{fig:LI-pulses}(d) obtained by displaying the output of the fast detector on a high bandwidth oscilloscope. The measured pulses (green solid line) match the simulated time dynamics taking into account the bandpass response of the detector (dashed black line). The measured pulse width is of about 400 ps, but since the effect of the bandpass mainly broadens the signal, the real pulse shape is narrower (dashed red line) with a FWHM of 190 ps. Last, the dependence on the modulation power of the light-curve curves is reported in Fig.\ref{fig:LI-pulses}(e). The trend clearly shows a reduction of lasing threshold and an increase of the slope efficiency and maximum output power as the modulation power is increased. 

\begin{figure}[htb!]
  \centering
  \includegraphics[width=.9\textwidth]{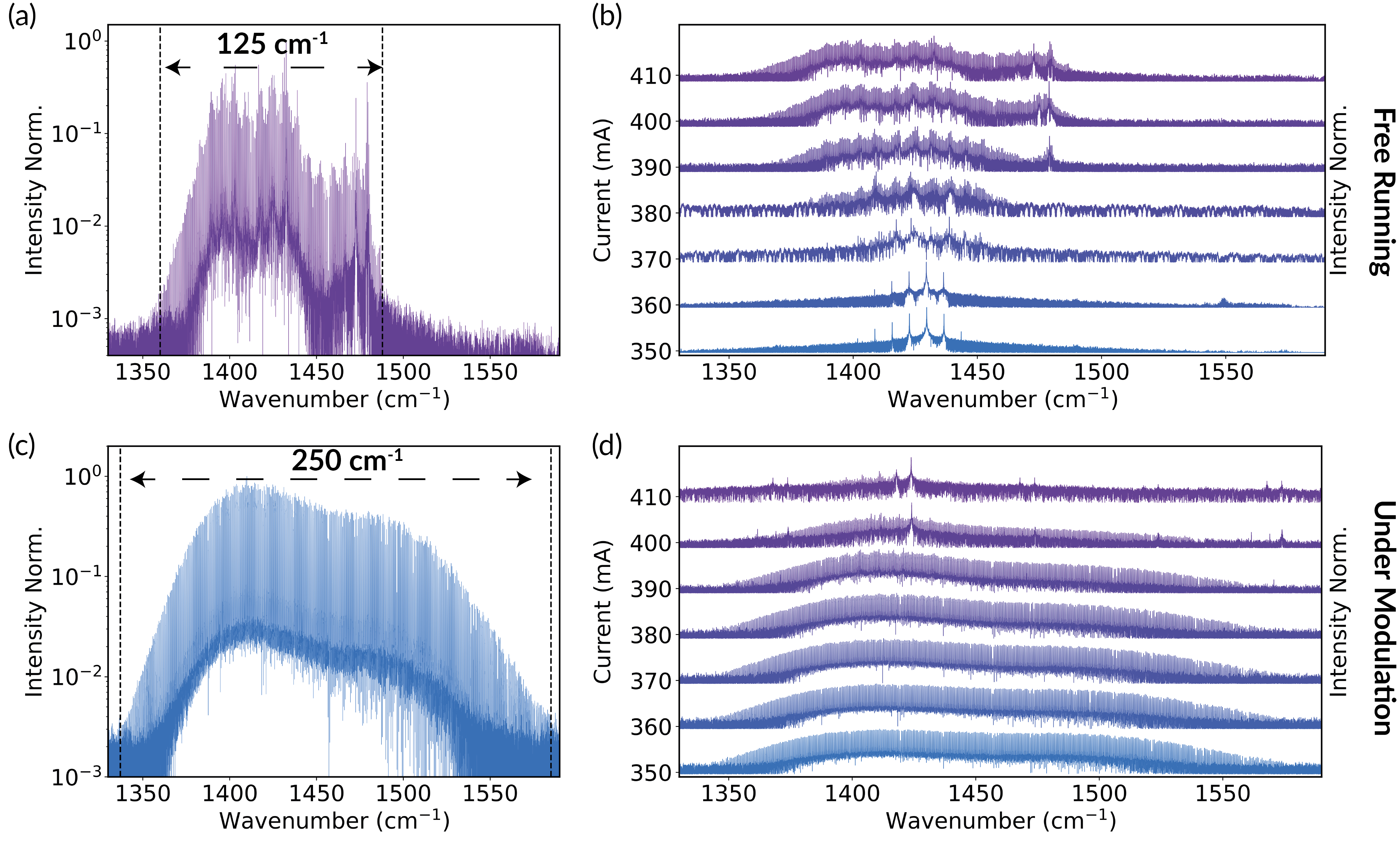}
  \caption{Spectra of the emitted light acquired through a high-resolution FTIR. \textbf{a)} Spectrum of the free running emission with a current bias of 410 mA \textbf{b)} Spectra of the free running emission under different current biases \textbf{c)} Spectrum of the emitted light under strong modulation ($f_M = 800$ MHz, $P_M = 29$ dBm) with a current bias of 350 mA \textbf{d)} Spectra of the emitted light under strong modulation ($f_M = 800$ MHz, $P_M = 29$ dBm) with different current biases.}
  \label{fig:spectra}
\end{figure}

These trends, together with the creation of short pulses, indicate that the device is operating in a regime very similar to the one of a gain switched device \cite{taschler_short_2023}. In particular, given a certain value of the DC current bias below CW lasing threshold, the current modulation will tend to create pulses as short as the time over which the gain overcomes the losses. If the bias current is too high, thermal effects and gain saturation will tend to reduce the effective gain therefore lowering the current at which roll-over happens. This kind of dynamic can also explain the measured spectral behavior. The typical emission in continuous wave (CW) operation, measured at a heat-sink temperature of 253 K, is reported in Fig.\ref{fig:spectra}(a) while the emission trend as a function of the bias current is reported in Fig.\ref{fig:spectra}(b). The spectrum develops from a few modes up to an irregular spectral shape covering a 125 cm$^{-1}$ bandwidth, with modes spaced by the repetition frequency, $f_{rep} \sim 15$ GHz. As expected, we observe a noisy behavior characteristic of unstable combs, in particular having an approximately regular mode spacing but with no signs of detectable electrical beatnotes. Instead, the spectral behavior under strong modulation is  consistent with the measured time dynamic and it is reported in  Fig.\ref{fig:spectra}(c-d). In particular in the latter, the spectra under a modulation frequency $f_M = 800$ MHz and a power $P_M =  29$  dBm are reported for different bias currents. When the bias current is about 350mA, the emission bandwidth is doubled with respect to the broadest free running spectrum and reaches 250 cm$^{-1}$ covering the full gain bandwidth (see also S\ref{SI:gain}). By increasing the bias current, the spectral shape is constant up to 390mA, while for higher values the envelope collapses in a few modes emission. In fact, the DC bias, given a fixed current modulation power, determines the residual intensity of the modes when the next current pulse arises which then impacts the strength of the effective FWM non-linearity. In particular, for low DC bias the modulation is able to turn on and off the device while for higher currents the device might not be completely switched off, therefore the photons in the cavity can experience longer timescale dynamics causing a collapse of the envelope and returning to a more unstable state as clearly shown in Fig.\ref{fig:spectra}(d). The emission behavior has also been characterized for different injection frequencies (see S\ref{SI:combs}).

\begin{figure}[htb!]
  \centering
  \includegraphics[width=.5\textwidth]{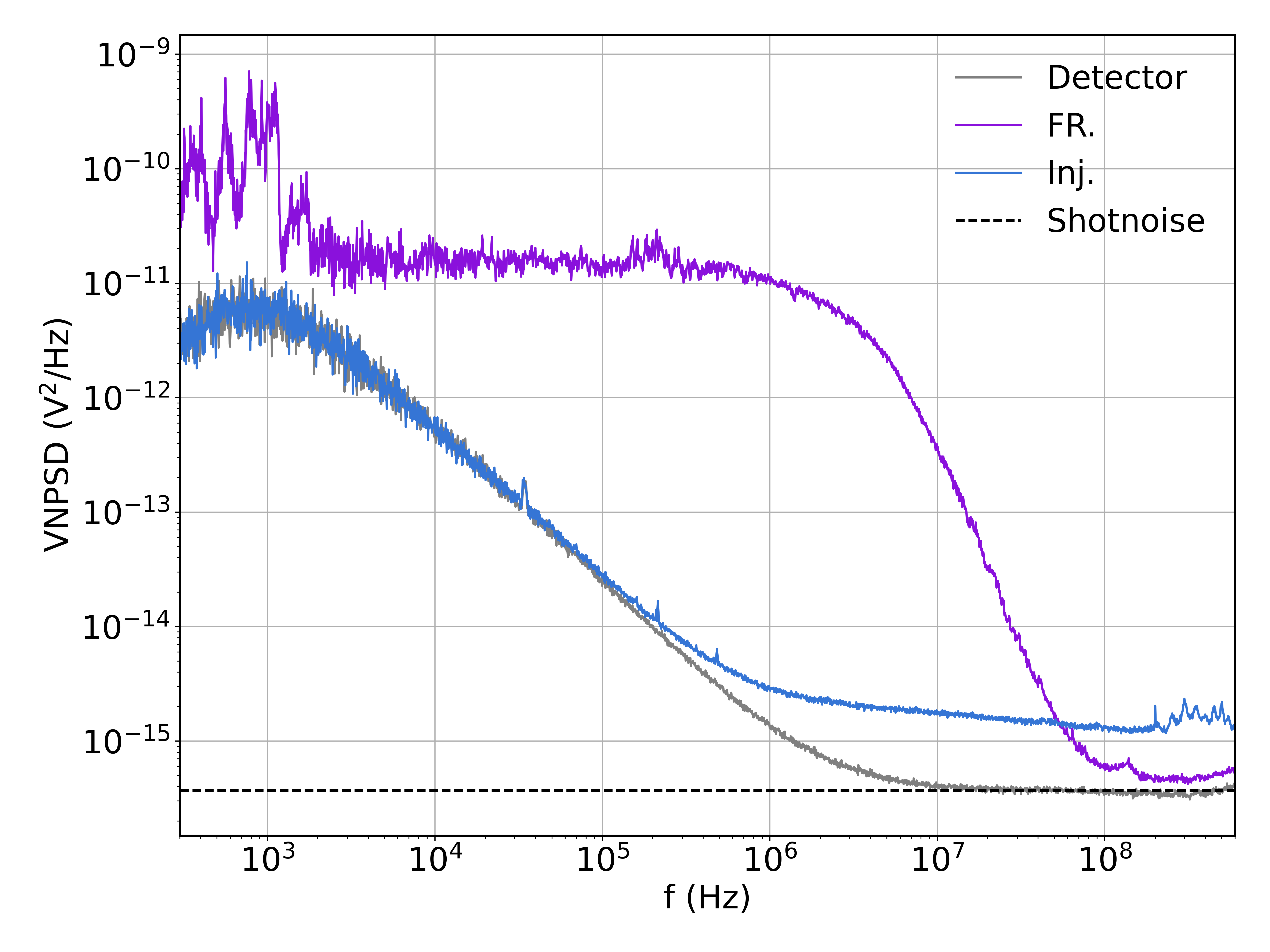}
  \caption{Voltage Noise Power Spectral Density (VNPSD) in the free running case with a bias current of 410mA (violet line) and in the strongly modulated case with a bias current of 350mA, $f_M = 800$ MHz, $P_M = 29$ dBm (blue line). The detector noise floor is also reported in gray, while the black dotted line is the computed shot noise for the free running case. The shot noise level in the modulated case would be a factor 3 lower than the one in the free running condition, so in the chosen scale the two curves would overlap and for this reason only the latter one is reported.}
  \label{fig:noise}
\end{figure}

The possibility to effectivelly suppress the four-wave mixing process on the short time scale dynamics allows to achieve a better amplitude noise figure, as presented in Fig.\ref{fig:noise}. The amplitude noise is acquired through a fast Peltier cooled HgCdTe (MCT) detector with a 1GHz nominal bandwidth, and it is reported in terms of Voltage Noise Power Spectral Density (VNPSD), both in the free running and modulated case in the operating condition where the maximum emission bandwidth is achieved. The noise in the free running case is relatively flat up to 1 MHz, showing a $1/f^2$ decay afterwards and reaching almost shot noise level above 100MHz. On the other hand, in the modulated case the VNPSD is reduced so much to be limited by the $1/f$ noise of the detector up to 1MHz where it reaches a plateau of about $1.5\cdot10^{-15}$ V$^2$/Hz, which is factor of 60 bigger than the expected shot noise level. The behavior at high frequency can be qualitatively explained by the fact that the fluctuations arising from the spontaneous emission are not damped by the gain clamping effect. Therefore, the system is reminiscent of an amplified spontaneous emission noise which contributes to a higher noise floor at higher frequency.

\begin{figure}[htb!]
  \centering
  \includegraphics[width=.9\textwidth]{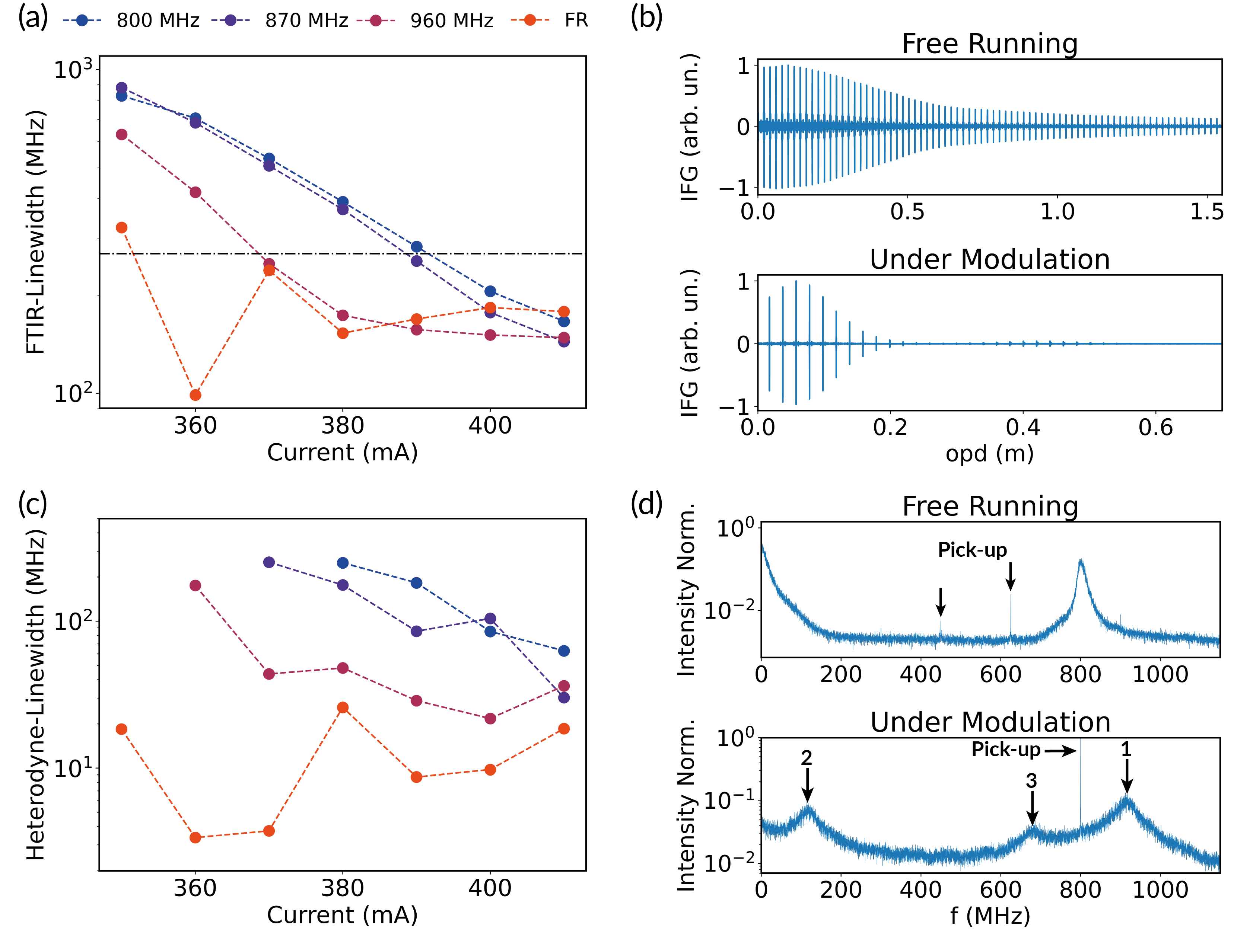}
  \caption{\textbf{a)} Linewidth of the emitted light estimated through FTIR measurements under different operation conditions as a function of the current bias for different modulation frequencies. The dashed horizontal line indicats the instrumental resolution of 270 MHz. \textbf{b)} Typical interferograms for the free running (410 mA, upper plot) and moduated case (350 mA, 800 MHz, lower plot). \textbf{c)}  Linewidth of the emitted light estimated through heterodyne measurements under different operation conditions as a function of the current bias for different modulation frequencies. \textbf{d)} Typical heterodyne measurements for the free running (410 mA, upper plot) and inject case (410 mA, 800 MHz, lower plot). Arrows indicate some Pick-up signal in both cases and the three peaks in the spectra of the modulated laser. }
  \label{fig:coherence}
\end{figure}

As expected, by operating the device under strong modulation, it is possible to reduce the amplitude noise and to prevent the onset of four-wave mixing. Since both aspects would concur to the generation of a  narrow linewidth it is expected that the latter is limited compared to the free running case. To address the general behavior, the laser linewidth was estimated through high resolution FTIR and heterodyne measurements. In the case of the FTIR measurements, a home made interferometer with a total optical path delay (OPD) of 1.6 m is used to retrieve the spectra shown in Fig.\ref{fig:spectra} and to extract the emission linewidth from the decaying envelope of the interferograms. In Fig.\ref{fig:coherence}(a) the linewidth dependence on the current bias for different injection frequencies is reported. In the free running condition, the linewidth tends to be quite constant and typically below the instrumental resolution (270 MHz), while for the modulated cases the typical linewidth is higher and spans from values of about 800 MHz down to the resolution of the FTIR. An example of measured interferograms in the free running and modulated case is reported in Fig.\ref{fig:coherence}(b), where the strong difference between the two different decaying envelopes can be quite clearly observed. Since some operating conditions produce linewidths below the instrumental resolution, an estimation through heterodyne measurements was necessary. The linewidth trend is displayed in Fig.\ref{fig:coherence}(c) and typical heterodyne beatings are reported in Fig.\ref{fig:coherence}(d). In the free running condition, the signal presents one well defined peak with a typical FWHM around 10 MHz and a strong low frequency noise which is in agreement with the trend shown in Fig.\ref{fig:noise}.

In the modulated case it is possible to distinguish three peaks, which correspond to the mixing of the local oscillator with the main emission peak and the two sidebands distanced by the modulation frequency, indicating that the laser is indeed in an amplitude modulation regime. Since we are acquiring with a sampling rate of 2.5Gs/s, we are able to distinguish two sidebands in the Nyquist limited bandwidth (principal at 900MHz, peak 1,  and sideband around 100MHz, peak 2), while the smallest peak around 700 MHz, peak 3, is the copy of the peak that should appear at 1.7GHz and it is folded back due to sub-sampling. Independently on the measurement technique, as the bias current increases the coherence improves and generally, this happens also for a fixed current and increasing injection frequency. The trend is compatible with the envelope and bandwidth trend, in particular all the regions at higher linewidth present a smooth emission envelope while the higher coherence points are the ones with the most irregular spectral shape since, as mentioned above, they are the ones experiencing the FWM non-linearity on a longer timescale. While the two measurement techniques show the same linewidth trend, they can give different absolute linewidth values. Specifically, heterodyne measurements often result in a lower linewidth value, consistent with the acquisition method. This difference arises because FTIR measurements integrate the linewidth over the entire mirror scan time (minutes), while heterodyne measurements have a shorter integration time (100$\mu$s). Consequently, the linewidth values differ due to technical noise limitations at this integration time.

\section*{Conclusions}
We proved that it is possible to exploit the full gain bandwidth of QCLs using strong RF modulation. If the bias current is sufficiently low to avoid gain saturation effects, the spectral envelope is extremely broad, smooth and the amplitude noise is limited by the detector noise below 1 MHz.  On the other hand, these new features are the result of a trade-off for lower coherence, with linewidths that can increase up to about 800MHz. This behavior can be explained by the fast switch on-off dynamic of the device which does not have time to build instabilities and it is corroborated by the time dynamics measurement which match quite well with the system simulated through a rate equation approach. In conclusion, this new way of operating QCLs which possibly have by design a very broadband gain, opens the possibility of using them as stable and bright sources for broadband and fast spectroscopy, resulting particularly appealing if used in combination with fast spectrometers \cite{markmann_frequency_2023} 

\newpage

\section*{Methods}

\subsection*{Strong Modulation}
The laser (SBCW3037, Alpes Lasers) is driven with a DC current source, ppq-Sense QubeCL. The RF signal is generate using a Rohde\& Schwarz SGS100A generator and amplified through an home-made high gain amplifier. The amplified RF signal is superimposed to the DC current through a bias tee, INSTOCK wireless components BT4120. 
\subsection*{Pulse Measurements}
The light emitted from the modulated device is directly shined on a fast photodetector, VIGO-UHSM-10.6, and the signal is acquired with a ultra-fast oscilloscope, HP-54120B.
\subsection*{Noise Measurements}
The light emitted from the device is directly shined on a fast photodetector, VIGO-UHSM-10.6. The signal is demodulated by a lock-in amplifier, Zurich Instruments UHFLI-600MHz, which records the amplitude noise by internally sweeping the local oscillator frequency.
\subsection*{Coherence Measurements}
The FTIR spectra are acquired though a home-made setup with a nominal optical path delay of 1.6m. The detector is a liquid nitrogen-cooled MCT detector, InfraRed Associates Inc. IRA-20-00113. The heteorodyne measurement is performed mixing an EC-QCL with one mode of the modulated laser selected through a spectral filter, and detecting the beating on a fast photodetector, VIGO-UHSM-10.6. The signal is aquired with a fast oscilloscope Teledyne Lecroy HDO6104-MS.

\section*{Acknowledgements}
The authors acknowledge  the use of ComponentLibrary (http://www.gwoptics.org/ComponentLibrary/) by Alexander Franzen used for realizing Fig.\ref{fig:setup}(a).

\section*{Funding}
The authors gratefully acknowledge founding from the ClosedLoop-LM project (SFA Advanced Manufactoring 2021-2024/ETH), from the MIRAQLS project (European Project 101070700) and from the Integrated, fast on-chip combs and detectors for THz photonics project (SNF$-$200021$-$213735). 
\newpage

\section*{Quantum Cascade Lasers as Broadband sources via Strong RF Modulation: Supplementary Material}

\renewcommand{\figurename}{S}

\counterwithout{figure}{section}
\counterwithout{equation}{section}
\setcounter{figure}{0}


\section*{Rate Equation Model}

The QCL is represented by a 3-level system (Fig.\ref{fig:LI-pulses}(a) of main text) \cite{faist_quantum_2013}, where the dynamics of the upper state population density $n_3$ (eq.\ref{eq:rate1}) is subject to an electrical pumping $J/e$. The current density $J$ can be constant ($J_0$) or time dependent in the case of strong modulation ($J_0 + J_M(t)$). The excited electrons can relax either non-radiatevely to the second or first level with rates $1/ \tau_{32},1 /\tau_{31}$ or radiatevely through stimulated emission with a depopulation term $Sg_c(n_3 - n_2)$, where $S$ is the photon flux density, $g_c$ is the gain cross-section and $n_2$ is the second level population density. The dynamics of the latter term is described by eq.\ref{eq:rate2} and takes into account the positive contribution from the radiative and non-radiative decay from level 3, while the depopulation term has a decay rate of $1/\tau_2$. The last contribution takes into account thermal back-filling by adding $n_2^{th} / \tau_2$, where $n_2^{th}$ is the thermal population density. Lastly, the photon flux dynamics is expressed in eq.\ref{eq:rate3} where, besides the stimulated emission term, we need to introduce the fraction of spontaneous emission $\beta$ into the lasing mode with a rate $1/\tau_3$ and the total losses $\alpha_{tot}$ (in our case mirror plus waveguide losses). The propagation of the light into a medium is taken into account by the term $c/n_r$ where $c$ is the speed of light and $n_r$ is the refractive index. The laser parameters are first estimated fitting the CW curve reported in Fig.\ref{fig:LI-pulses}(b) and are reported in Tab.\ref{tab:rate}. 

\begin{equation}
    \frac{dn_3}{dt} = \frac{J_0 + J_M(t)}{e}  - \frac{n_3}{\tau_{32}} - \frac{n_3}{\tau_{31}} - Sg_c(n_3 - n_2)
\label{eq:rate1}
\end{equation}
\begin{equation}
    \frac{dn_2}{dt} = \frac{n_3}{\tau_{32}} + Sg_c(n_3-n_2) - \frac{n_2 - n_2^{th}}{\tau_2}
\label{eq:rate2}
\end{equation}
\begin{equation}
    \frac{dS}{dt} = \frac{c}{n_r}\bigg( Sg_c(n_3- n_2) - S\alpha_{tot} + \beta\frac{n_3}{\tau_{sp}} \bigg)
\label{eq:rate3}
\end{equation}

\begin{table}[htb!]
\centering
\begin{tabular}{ l c c c}
\textbf{Rate Equation Parameters} & & \\
\hline
Upper-lower level lifetime & $\tau_{31}$ & 0.9 & ps \\ 
Upper-ground level lifetime & $\tau_{32}$ & 1.1 & ps \\  
Lower level lifetime & $\tau_2$ & 0.55 & ps \\
Spontaneous emission lifetime & $\tau_{sp}$& 150 & ns\\
Gain cross section & $g_c$ & 6.5$\cdot10^{-9}$ & cm   \\
Thermal population & $n_2^{th}$ & 1.25$\cdot 10^9$ & cm$^{-2}$ \\
Refractive index & $n_r$& 3.3 & \\
Total losses & $\alpha_{tot}$ & 6.17 & cm$^{-1}$ \\
Sp. Emitted light into lasing mode & $\beta$& 1$\cdot10^{-3}$ & \\
\hline
Number of QCL period & $N_p$ & 40 &\\
Laser length & $L$ & 3 & mm\\
Laser width & $w$& 8 & $\mu$m \\
Wavelength &$\lambda$ & 7 &  $\mu$m\\
\hline
\end{tabular}
\caption{Rate equation parameters (upper part) and laser parameters (lower part) used to fit the light-current curves and the time dependent laser behavior.}
\label{tab:rate}
\end{table}

As mentioned in the main text, the rate equation model can predict correctly the time dynamics of the laser and the light-current characteristics, under the assumption that we can neglect gain saturation and thermal effects. The first condition is ensured for operation points that are far away from the roll-over, while the second one can not be disregarded completely. In particular, the heating of the device will give rise to lower output powers and lower slope efficiencies. For this reason, it is necessary to introduce a scaling factor $\sigma$ that corrects the simulated power to match the measured one. This parameter empirically takes into account the heating effects and therefore is a measure of the thermal behavior under strong modulation. This operation does not effect the pulse width estimation (see Fig.\ref{fig:LI-pulses}(d)) since it only acts on the total emitter power.

\begin{figure}[htb!]
  \centering
  \includegraphics[width=.9\textwidth]{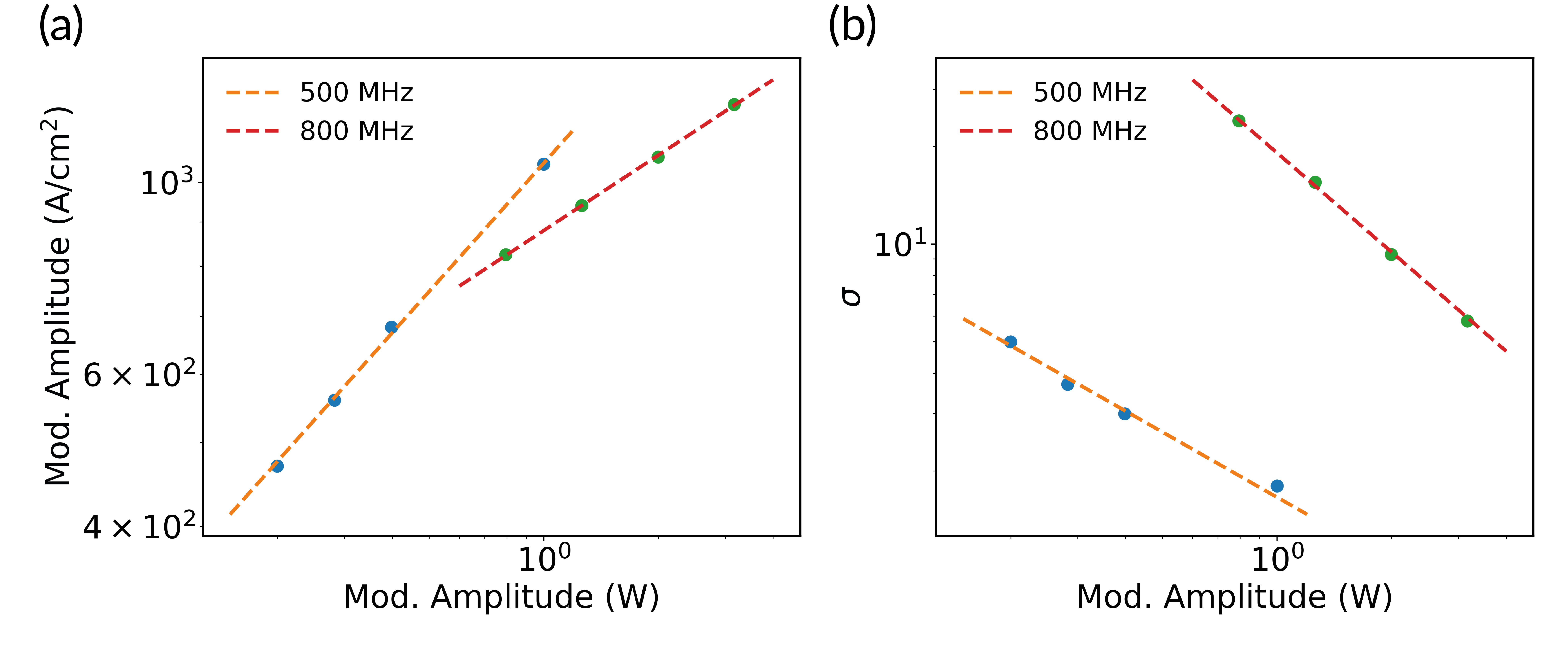}
  \caption{\textbf{a)} Modulation amplitude as a function of the modulation power for modulation frequencies of 500 MHz and 800 MHz. \textbf{b)} Scaling parameter $\sigma$ as a function of the modulation power for modulation frequencies of 500 MHz and 800 MHz. In both cases the points represent the extracted values from the rate equation model, while the dotted lines are the best fit using a power law.}
  \label{SI:scaling}
\end{figure}

in S\ref{SI:scaling}, we report the power dependence of the current density modulation amplitude and of the scaling factor, extracted by matching the light-current curves for two modulation frequencies. The modulation amplitude is expected to scale with the square root of the modulation power, and indeed the slope $a$ of the data referred to 500 MHz extracted through a best fit with a power law gives a value of $a = 0.49 \pm 0.01 $ kA/Wcm$^-2$. Instead, the slope for the 800 MHz case results in $a = 0.289 \pm 0.004$ kA/Wcm$^-2$ which can be simply explained by the fact that at higher frequencies the impedance matching of the QCL becomes worse \cite{kapsalidis_mid-infrared_2021} and the nominal modulation power is not totally converted into modulation current. 

The scaling parameter also follows a power law, with slopes $a = -0.68 \pm 0.06$  kA/Wcm$^-2$ and $a = -1.02 \pm 0.03$  kA/Wcm$^-2$ for 500 MHz and 800 MHz modulation frequencies, respectively. These values are also in agreement with the expectations. In particular, the lower frequency modulation creates pulses that on average are farther apart, allowing the device to cool down and effectively working with a lower duty cycle. On the contrary, at higher frequencies the duty cycle is higher. Therefore, heat dissipation is worse, resulting a generally higher correction factor slope.

\newpage

\section*{Modal Gain Estimation}
\begin{figure}[htb!]
  \centering
  \includegraphics[width=\textwidth]{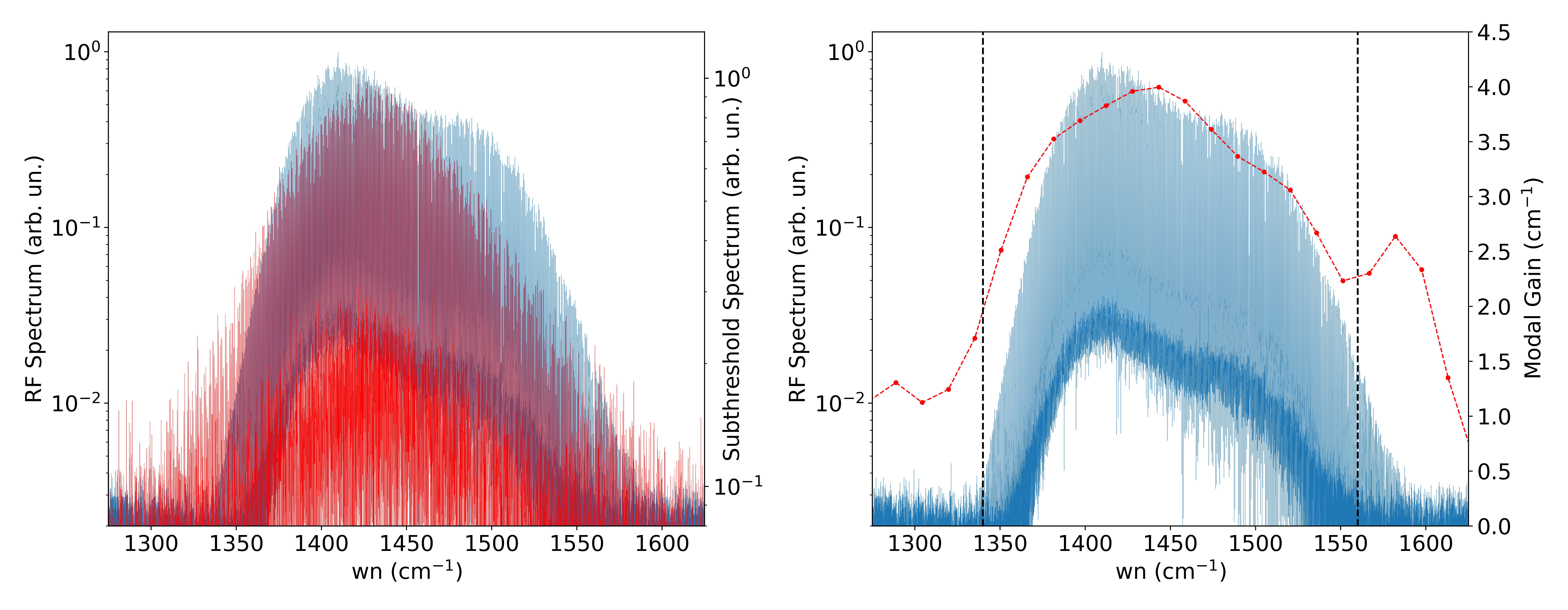}
  \caption{Comparison between the emission spectrum of the device under strong injection (350 mA, 800 MHz), the sub-threshold luminescence (on the left), and the gain profile (on the right). The dashed vertical lines delimit the region where the SNR is high enough to have a reasonable estimation of the the modal gain.}
  \label{SI:gain}
\end{figure}

\section*{Laser Spectra at Different Modulation Frequencies}

\begin{figure}[htb!]
  \centering
  \includegraphics[width=\textwidth]{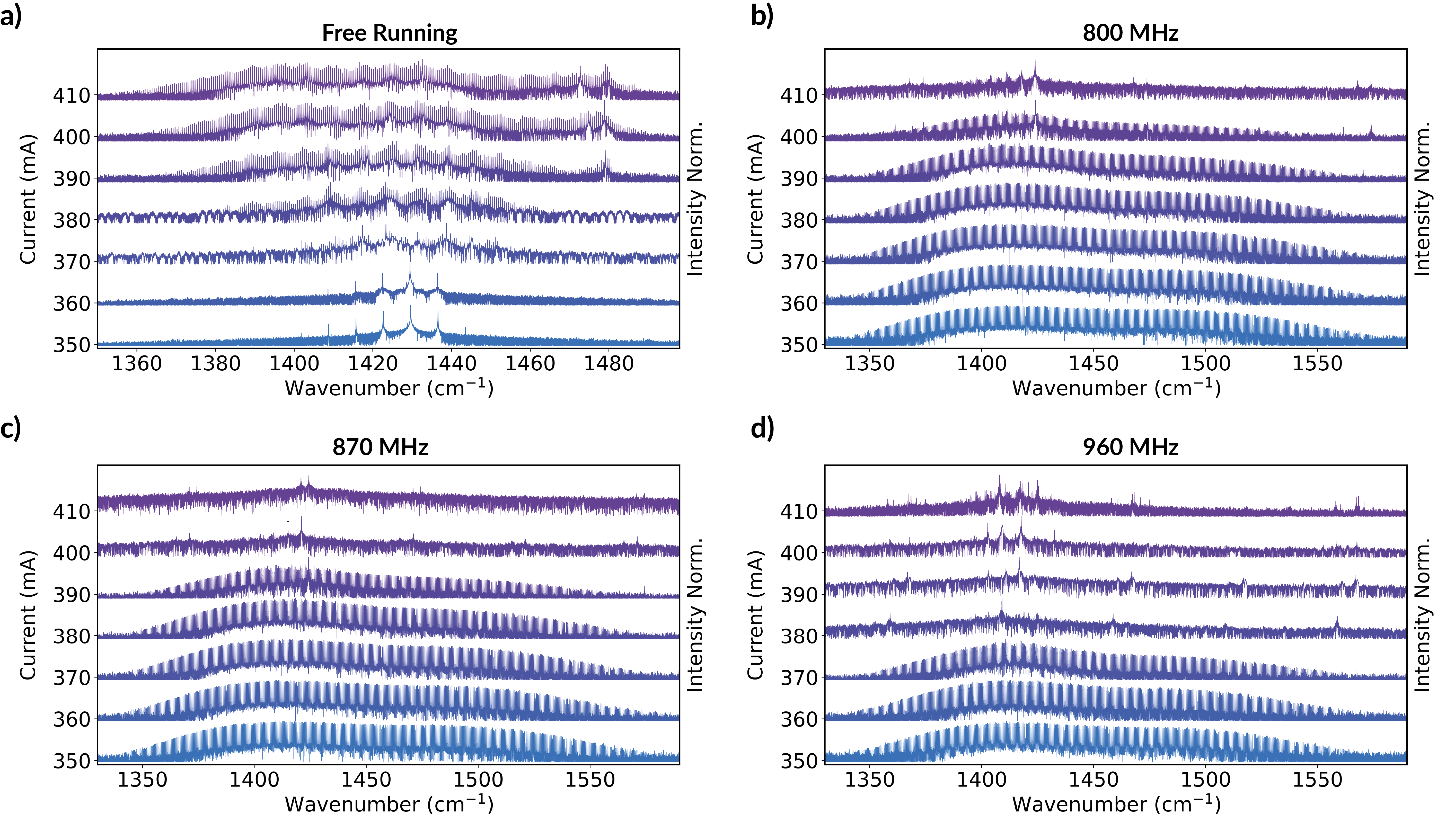}
  \caption{Spectral behavior of the device under different operating conditions as a function of the current bias. The free running condition, and the strongly modulated cases, are reported in the following order: \textbf{a)} Free Running, \textbf{b)} 800 MHz, \textbf{c)} 870 MHz, \textbf{d)} 960 MHz. The RF power for all the strongly modulated cases is 29dBm.}
  \label{SI:combs}
\end{figure}

The different spectral behavior of the laser under strong modulation as a function of the modulation frequency is reported in Fig.\ref{SI:combs} and it is simply explained by an impedance matching problem. As mentioned in the rate equation section, for a fixed modulation power (29 dBm for the spectra under consideration) the effective current modulation becomes less efficient for increasing frequency  \cite{kapsalidis_mid-infrared_2021}. Since the regular spectral envelope and broadband emission are guaranteed whenever we are able to switch on and off the laser, it is possible that this condition is not guaranteed for a fixed current bias at different modulation frequencies. For example, the 390 mA current point produces a quite regular envelope at $f_M = 800$ MHz, while for $f_M = 870$ MHz the spectrum starts being less regular until for $f_M = 960$ MHz the spectrum is collapsed, indicating that the effective modulation current decreases for increasing frequency since at some point it is not possible to switch the device on and off completely anymore and the light can start experiencing longer timescales in the cavity, therefore producing an irregular envelope. In conclusion, the spectral frequency dependence directly translates into the effective current modulation which might not be high enough for a complete switch off of the device. In addition, higher modulation frequency generally implies a higher emission duty-cycle directly impacting the thermal performance of the device.

\newpage

\bibliographystyle{biolett}
\bibliography{references}

\begin{thebibliography}{10}
\expandafter\ifx\csname urlstyle\endcsname\relax
  \providecommand{\doi}[1]{doi:\discretionary{}{}{}#1}\else
  \providecommand{\doi}{doi:\discretionary{}{}{}\begingroup \urlstyle{rm}\Url}\fi

\bibitem{willer_near-_2006}
Willer U, Saraji M, Khorsandi A, Geiser P, Schade W. 2006 Near- and mid-infrared laser monitoring of industrial processes, environment and security applications.
\newblock \emph{Optics and Lasers in Engineering} \textbf{44}, 7, 699--710.
\newblock (\doi{10.1016/j.optlaseng.2005.04.015}).

\bibitem{wang_application_2008}
Wang L, Mizaikoff B. 2008 Application of multivariate data-analysis techniques to biomedical diagnostics based on mid-infrared spectroscopy.
\newblock \emph{Analytical and Bioanalytical Chemistry} \textbf{391}, 5, 1641--1654.
\newblock (\doi{10.1007/s00216-008-1989-9}).

\bibitem{jung_next-generation_2017}
Jung D, Bank S, Lee ML, Wasserman D. 2017 Next-generation mid-infrared sources.
\newblock \emph{Journal of Optics} \textbf{19}, 12, 123001.
\newblock (\doi{10.1088/2040-8986/aa939b}).

\bibitem{ricker_broadband_2017}
Ricker RJ, Provence SR, Norton DT, Boggess TF, Prineas JP. 2017 Broadband mid-infrared superlattice light-emitting diodes.
\newblock \emph{Journal of Applied Physics} \textbf{121}, 18, 185701.
\newblock (\doi{10.1063/1.4983023}).

\bibitem{yu_experimental_2016}
Yu Y, Gai X, Ma P, Vu K, Yang Z, Wang R, Choi DY, Madden S, Luther-Davies B. 2016 Experimental demonstration of linearly polarized 2–10  μm supercontinuum generation in a chalcogenide rib waveguide.
\newblock \emph{Optics Letters} \textbf{41}, 5, 958--961.
\newblock (\doi{10.1364/OL.41.000958}).
\newblock Publisher: Optica Publishing Group.

\bibitem{zorin_advances_2022}
Zorin I, Gattinger P, Ebner A, Brandstetter M. 2022 Advances in mid-infrared spectroscopy enabled by supercontinuum laser sources.
\newblock \emph{Optics Express} \textbf{30}, 4, 5222--5254.
\newblock (\doi{10.1364/OE.447269}).
\newblock Publisher: Optica Publishing Group.

\bibitem{corwin_fundamental_2003}
Corwin KL, Newbury NR, Dudley JM, Coen S, Diddams SA, Weber K, Windeler RS. 2003 Fundamental {Noise} {Limitations} to {Supercontinuum} {Generation} in {Microstructure} {Fiber}.
\newblock \emph{Physical Review Letters} \textbf{90}, 11, 113904.
\newblock (\doi{10.1103/PhysRevLett.90.113904}).

\bibitem{klimczak_direct_2016}
Klimczak M, Soboń G, Kasztelanic R, Abramski KM, Buczyński R. 2016 Direct comparison of shot-to-shot noise performance of all normal dispersion and anomalous dispersion supercontinuum pumped with sub-picosecond pulse fiber-based laser.
\newblock \emph{Scientific Reports} \textbf{6}, 1, 19284.
\newblock (\doi{10.1038/srep19284}).
\newblock Number: 1 Publisher: Nature Publishing Group.

\bibitem{faist_quantum_1994}
Faist J, Capasso F, Sivco DL, Sirtori C, Hutchinson AL, Cho AY. 1994 Quantum {Cascade} {Laser}.
\newblock \emph{Science} \textbf{264}, 5158, 553--556.
\newblock (\doi{10.1126/science.264.5158.553}).
\newblock \_eprint: https://www.science.org/doi/pdf/10.1126/science.264.5158.553.

\bibitem{lin_type-ii_nodate}
Lin CH, Yang RQ, Zhang D, Murry SJ, Pei SS, Allerman AA, Kurtz SR Type-{II} interband quantum cascade laser at 3.8 µm .

\bibitem{hugi_mid-infrared_2012}
Hugi A, Villares G, Blaser S, Liu HC, Faist J. 2012 Mid-infrared frequency comb based on a quantum cascade laser.
\newblock \emph{Nature} \textbf{492}, 7428, 229--233.
\newblock (\doi{10.1038/nature11620}).

\bibitem{faist_quantum_2016}
Faist J, Villares G, Scalari G, Rösch M, Bonzon C, Hugi A, Beck M. 2016 Quantum {Cascade} {Laser} {Frequency} {Combs}.
\newblock \emph{Nanophotonics} \textbf{5}, 2, 272--291.
\newblock (\doi{10.1515/nanoph-2016-0015}).

\bibitem{sterczewski_interband_2021}
Sterczewski LA, Bagheri M, Frez C, Canedy CL, Vurgaftman I, Kim M, Kim CS, Merritt CD, Bewley WW, Meyer JR. 2021 Interband cascade laser frequency combs.
\newblock \emph{Journal of Physics: Photonics} \textbf{3}, 4, 042003.
\newblock (\doi{10.1088/2515-7647/ac1ef3}).

\bibitem{villares_dual-comb_2014}
Villares G, Hugi A, Blaser S, Faist J. 2014 Dual-comb spectroscopy based on quantum-cascade-laser frequency combs.
\newblock \emph{Nature Communications} \textbf{5}, 1, 5192.
\newblock (\doi{10.1038/ncomms6192}).

\bibitem{gianella_high-resolution_2020}
Gianella M, Nataraj A, Tuzson B, Jouy P, Kapsalidis F, Beck M, Mangold M, Hugi A, Faist J, Emmenegger L. 2020 High-resolution and gapless dual comb spectroscopy with current-tuned quantum cascade lasers.
\newblock \emph{Optics Express} \textbf{28}, 5, 6197.
\newblock (\doi{10.1364/OE.379790}).

\bibitem{sterczewski_mid-infrared_2020}
Sterczewski LA, Bagheri M, Frez C, Canedy CL, Vurgaftman I, Meyer JR. 2020 Mid-infrared dual-comb spectroscopy with room-temperature bi-functional interband cascade lasers and detectors.
\newblock \emph{Applied Physics Letters} \textbf{116}, 14, 141102.
\newblock (\doi{10.1063/1.5143954}).

\bibitem{feng_passively_2020}
Feng T, Shterengas L, Hosoda T, Kipshidze G, Belyanin A, Teng C, Westberg J, Wysocki G, Belenky G. 2020 Passively {Mode}-{Locked} 2.7 and 3.2 μm {GaSb}-{Based} {Cascade} {Diode} {Lasers}.
\newblock \emph{Journal of Lightwave Technology} \textbf{38}, 7, 1895--1899.
\newblock (\doi{10.1109/JLT.2020.2971605}).
\newblock Conference Name: Journal of Lightwave Technology.

\bibitem{singleton_evidence_2018}
Singleton M, Jouy P, Beck M, Faist J. 2018 Evidence of linear chirp in mid-infrared quantum cascade lasers.
\newblock \emph{Optica} \textbf{5}, 8, 948.
\newblock (\doi{10.1364/OPTICA.5.000948}).

\bibitem{heckelmann_quantum_2023}
Heckelmann I, Bertrand M, Dikopoltsev A, Beck M, Scalari G, Faist J. 2023 Quantum walk comb in a fast gain laser.
\newblock \emph{Science} \textbf{382}, 6669, 434--438.
\newblock (\doi{10.1126/science.adj3858}).
\newblock Publisher: American Association for the Advancement of Science.

\bibitem{gmachl_ultra-broadband_2002}
Gmachl C, Sivco DL, Colombelli R, Capasso F, Cho AY. 2002 Ultra-broadband semiconductor laser.
\newblock \emph{Nature} \textbf{415}, 6874, 883--887.
\newblock (\doi{10.1038/415883a}).
\newblock Number: 6874 Publisher: Nature Publishing Group.

\bibitem{bandyopadhyay_high_2014}
Bandyopadhyay N, Bai Y, Slivken S, Razeghi M. 2014 High power operation of λ ∼ 5.2–11 \textit{μ} m strain balanced quantum cascade lasers based on the same material composition.
\newblock \emph{Applied Physics Letters} \textbf{105}, 7, 071106.
\newblock (\doi{10.1063/1.4893746}).

\bibitem{hugi_external_2010}
Hugi A, Maulini R, Faist J. 2010 External cavity quantum cascade laser.
\newblock \emph{Semiconductor Science and Technology} \textbf{25}, 8, 083001.
\newblock (\doi{10.1088/0268-1242/25/8/083001}).

\bibitem{villares_quantum_2015}
Villares G, Faist J. 2015 Quantum cascade laser combs: effects of modulation and dispersion.
\newblock \emph{Optics Express} \textbf{23}, 2, 1651.
\newblock (\doi{10.1364/OE.23.001651}).

\bibitem{opacak_theory_2019}
Opačak N, Schwarz B. 2019 Theory of {Frequency}-{Modulated} {Combs} in {Lasers} with {Spatial} {Hole} {Burning}, {Dispersion}, and {Kerr} {Nonlinearity}.
\newblock \emph{Physical Review Letters} \textbf{123}, 24, 243902.
\newblock (\doi{10.1103/PhysRevLett.123.243902}).

\bibitem{beiser_engineering_2021}
Beiser M, Opačak N, Hillbrand J, Strasser G, Schwarz B. 2021 Engineering the spectral bandwidth of quantum cascade laser frequency combs.
\newblock \emph{Optics Letters} \textbf{46}, 14, 3416.
\newblock (\doi{10.1364/OL.424164}).

\bibitem{singleton_combs_2021}
Singleton M. 2021 \emph{Combs in {Quantum} {Cascade} {Lasers}: {Linear} {Chirp} and {Pulse} {Compression}}.
\newblock Doctoral {Thesis}, ETH Zurich, Zurich.
\newblock (\doi{10.3929/ethz-b-000522464}).

\bibitem{khurgin_coherent_2014}
Khurgin JB, Dikmelik Y, Hugi A, Faist J. 2014 Coherent frequency combs produced by self frequency modulation in quantum cascade lasers.
\newblock \emph{Applied Physics Letters} \textbf{104}, 8, 081118.
\newblock (\doi{10.1063/1.4866868}).

\bibitem{hillbrand_coherent_2019}
Hillbrand J, Andrews AM, Detz H, Strasser G, Schwarz B. 2019 Coherent injection locking of quantum cascade laser frequency combs.
\newblock \emph{Nature Photonics} \textbf{13}, 2, 101--104.
\newblock (\doi{10.1038/s41566-018-0320-3}).

\bibitem{schneider_controlling_2021}
Schneider B, Kapsalidis F, Bertrand M, Singleton M, Hillbrand J, Beck M, Faist J. 2021 Controlling {Quantum} {Cascade} {Laser} {Optical} {Frequency} {Combs} through {Microwave} {Injection}.
\newblock \emph{Laser \& Photonics Reviews} \textbf{15}, 12, 2100242.
\newblock (\doi{10.1002/lpor.202100242}).

\bibitem{st-jean_injection_2014}
St-Jean MR, Amanti MI, Bernard A, Calvar A, Bismuto A, Gini E, Beck M, Faist J, Liu HC, Sirtori C. 2014 Injection locking of mid-infrared quantum cascade laser at 14 {GHz}, by direct microwave modulation.
\newblock \emph{Laser \& Photonics Reviews} \textbf{8}, 3, 443--449.
\newblock (\doi{10.1002/lpor.201300189}).
\newblock \_eprint: https://onlinelibrary.wiley.com/doi/pdf/10.1002/lpor.201300189.

\bibitem{hugi_external_2009}
Hugi A, Terazzi R, Bonetti Y, Wittmann A, Fischer M, Beck M, Faist J, Gini E. 2009 External cavity quantum cascade laser tunable from 7.6 to 11.4 μm.
\newblock \emph{Applied Physics Letters} \textbf{95}, 6, 061103.
\newblock (\doi{10.1063/1.3193539}).

\bibitem{bidaux_measurements_2015}
Bidaux Y, Terazzi R, Bismuto A, Gresch T, Blaser S, Muller A, Faist J. 2015 Measurements and simulations of the optical gain and anti-reflection coating modal reflectivity in quantum cascade lasers with multiple active region stacks.
\newblock \emph{Journal of Applied Physics} \textbf{118}, 9, 093101.
\newblock (\doi{10.1063/1.4929810}).

\bibitem{beck_continuous_2002}
Beck M, Hofstetter D, Aellen T, Faist J, Oesterle U, Ilegems M, Gini E, Melchior H. 2002 Continuous {Wave} {Operation} of a {Mid}-{Infrared} {Semiconductor} {Laser} at {Room} {Temperature}.
\newblock \emph{Science} \textbf{295}, 5553, 301--305.
\newblock (\doi{10.1126/science.1066408}).
\newblock Publisher: American Association for the Advancement of Science.

\bibitem{hofstetter_measurement_1999}
Hofstetter D, Faist J. 1999 Measurement of semiconductor laser gain and dispersion curves utilizing {Fourier} transforms of the emission spectra.
\newblock \emph{IEEE Photonics Technology Letters} \textbf{11}, 11, 1372--1374.
\newblock (\doi{10.1109/68.803049}).
\newblock Conference Name: IEEE Photonics Technology Letters.

\bibitem{faist_quantum_2013}
Faist J. 2013 \emph{Quantum {Cascade} {Lasers}}.
\newblock Oxford University Press.
\newblock (\doi{10.1093/acprof:oso/9780198528241.001.0001}).

\bibitem{taschler_short_2023}
Täschler P, Miller L, Kapsalidis F, Beck M, Faist J. 2023 Short pulses from a gain-switched quantum cascade laser.
\newblock \emph{Optica} \textbf{10}, 4, 507--512.
\newblock (\doi{10.1364/OPTICA.485407}).
\newblock Publisher: Optica Publishing Group.

\bibitem{markmann_frequency_2023}
Markmann S, Franckié M, Bertrand M, Shahmohammadi M, Forrer A, Jouy P, Beck M, Faist J, Scalari G. 2023 Frequency chirped {Fourier}-{Transform} spectroscopy.
\newblock \emph{Communications Physics} \textbf{6}, 1, 1--12.
\newblock (\doi{10.1038/s42005-023-01157-5}).
\newblock Number: 1 Publisher: Nature Publishing Group.

\bibitem{kapsalidis_mid-infrared_2021}
Kapsalidis F, Schneider B, Singleton M, Bertrand M, Gini E, Beck M, Faist J. 2021 Mid-infrared quantum cascade laser frequency combs with a microstrip-like line waveguide geometry.
\newblock \emph{Applied Physics Letters} \textbf{118}, 7, 071101.
\newblock (\doi{10.1063/5.0040882}).

\end{thebibliography}

\end{document}